# Half-Heusler Compounds: Promising Materials For Mid-To-High Temperature Thermoelectric Conversion

**S. Joseph Poon**

**Department of Physics, University of Virginia, Charlottesville, VA 22904-4714, USA**

**Abstract**

Half-Heusler compounds (space group $Fm\bar{3}m$) have garnered increasing attention in recent years in the thermoelectric community. Three decades ago, refractory RNiSn half-Heusler compounds (R represents refractory metals such as Hf, Zr, Ti) were found to be narrow-gap semiconductors with large Seebeck coefficients in 100s of micro-volt per Kelvin. Today, half-Heusler (HH) compounds have emerged as promising thermoelectric materials in the intermediate temperature range (400-800°C). HH materials are endowed with good thermal stability and scalability. Thermoelectric n-p modules based on HH materials demonstrate conversion efficiency near 10% and power density output near 9 W/cm$^2$. The objective of this article is to present a historical account of the research and development of thermoelectric half-Heusler compounds. Particularly, there have been notable achievements since 2012 thanks to the emergence of new approaches. As a result, ZT has risen from ~1 to 1.5. The various advances made since the early 1990s to the present are recounted by categorizing half-Heusler materials into three generations (Gen): Gen-1 Gen-2, and Gen-3 HH materials.

# 1. Introduction

More than half of the global energy produced from available sources is wasted, mostly in the form of heat[1]. The loss of useful energy amounts to a significant portion of the global GDP. To date, converting waste heat into useful energy is a major challenge due to low efficiency and high cost. The opportunities for waste heat recovery can be categorized according to the temperature ranges of the waste heat sources into the following areas[2]: (i) low-grade heat harvesting (450°F/232°C and lower) from, e.g., power plants, buildings, household utilities, computing centers, sea water, and geothermal sources; (ii) medium-grade (450°F/232°C to 1,200°F/650°C) to high–grade (1,200°F/650°C and higher) heat harvesting from, e.g., vehicle and jet engines as well as industrial plants. An energy harvest device based on thermoelectric (TE) materials can convert waste heat directly into electricity[3]. The device is all solid state, has no moving parts, and is portable. TE conversion technology is among the simplest direct energy conversion technologies, while leaving a minimal negative impact on the environment. TE devices are portable and can conveniently be integrated with other energy-conversion technologies. Critical to the TE devices development is the provision of high-performance thermoelectric materials.

A class of thermoelectric materials known as half-Heusler compounds (space group $Fm\bar{3}m$) have gained much attention in recent years in the thermoelectric community. In particular, the RNiSn-type half-Heusler compounds (R=Hf, Zr, Ti) were first reported to be narrow-gap semiconductors by Aliev and co-workers three decades ago[4,5,6]. These authors measured a bandgap of about 0.2 eV in optical and electrical transport experiments. They also reported a large n-type Seebeck coefficient on the order of a couple hundred micro-volts per degree Kelvin. The finding of large thermopower in RNiSn compounds apparently did not garner much attention until the mid-1990s when Cook et al[7] and Kloc et al[8] reported large thermopowers in excess of -300 μV/K in some well-crystallized RNiSn samples. Investigation of the origin of bandgap in these metal-based semiconductors was performed by Ogut and Rabe in 1995 [9]. The latter authors identified the important role of pd hybridization and dd interaction in forming the bandgap.

The goal of this article is to review the research and development of thermoelectric half-Heusler (HH) compounds in the last three decades. In particular, the last decade has seen a rapid development of half-Heusler compounds thanks to the implementation of new synthesis methods and interdisciplinary approaches. Particularly since 2013, the dimensionless figure of merit, ZT, an important measure of thermoelectric performance, has risen rapidly from 1 to ~1.5. The advances made in different stages of development are remarkable. Accordingly, we have structured this review chronologically around three categories of thermoelectric half-Heusler materials, hereafter referred to as Gen-1 Gen-2, and Gen-3 HH materials (Gen is an abbreviation for Generation).

# 2. Background

## 2.1 Basics of thermoelectric figure of merit

The thermoelectric dimensionless figure of merit, ZT, is defined as $ZT=(S^2\sigma/\kappa)T$, where S is the Seebeck coefficient, σ is the electrical conductivity, and κ is the thermal conductivity[3]. κ is the sum of $\kappa_e$ and $\kappa_L$, electronic contribution and lattice contribution to κ, respectively. $\kappa_e$ is in turn related to σ via the expression $\kappa_e = L\sigma T$, where $L$ is the Lorenz number. $S^2\sigma$ is the power factor (PF). Some authors defined PF as $S^2\sigma T$, in which case, PF has the same dimension as κ (in W/m-K). For clarification purpose, we will use PF*T, the power factor temperature product, in the latter definition. Alternatively, ZT can be written as follows:

$$ZT = \frac{S^2/L}{\kappa_L/(L\sigma T)+1} \quad (1)$$

Assuming a single carrier-type model (e.g., n-type), then $\sigma = nq\mu_c$, and $S \sim m_d^* T/n^{2/3}$ as approximated using the Mott's formula[10], where $\mu_c$ is the carrier mobility, and $m_d^*$ is the total density of states (DOS). Direct substitution of these microscopic electronic bandstructure and transport parameters in equation (1) reveals that in order to enhance ZT, the parameters $m_d^*$, $\mu_c$, and T must be increased while $\kappa_L$ must be decreased. These requirements are encapsulated in the quality factor known as B factor[11] expressed as follows:

$$B \sim \mu_c m_d^{*3/2} T^{5/2}/\kappa_L \quad (2)$$

Equation (2) infers that the electronic and lattice properties can be independently controlled to increase B and thus ZT. However, the real situation is more complex. This is because $\mu_c$ also depends on the effective mass, as follows:

$$\mu_c \sim \frac{1}{m_c^* m_b^{*3/2} \varepsilon^2} \quad (3)$$

where $m_c^*$ and $m_b^*$ are the conductivity effective mass and single-valley DOS effective mass, respectively. The effective masses mentioned are related to each other. In fact, $m_d^*$ is given by $N_V^{2/3} m_b^*$, where $N_V$ is the valley degeneracy. The scaling relation in equation (2) can be rewritten as $B \sim N_V T^{5/2}/m_c^* \varepsilon^2 \kappa_L$. The band degeneracy therefore stands out as an important electronic factor for improving TE performance. ε is the acoustic deformation potential for carrier scattering from phonons. A large ε would reduce $\mu_c$, while also lowering $\kappa_L$. The latter can also be reduced by other scattering mechanisms. As a result, one can strategically design composition and microstructure to separately control the electronic and phonon transport properties.

*2.2 Half-Heusler compounds, crystal structure and composition*

Heusler compounds are among the best known intermetallic compounds. The compositions of Heusler compounds are represented by the chemical formula $XY_2Z$ for full-Heusler compounds and XYZ for half-Heusler compounds, respectively, where X and Y are metals, and Z is a main group element. Stable Heusler compounds can form in many compositions[12, 13, 14, 15, 16, 17, 18]. For most of the Heusler compounds studied to date, X is either a transition metal, a lanthanide metal, or a noble metal, and Y is a transition metal or noble metal. The crystal structure of the full-Heusler (FH) compounds is of the $L2_1$ type (space group $Fm\bar{3}m$) consisting of four interpenetrating face-centered cubic (fcc) sublattices, each fully occupied by atoms. If one of the two Y sublattices is vacant, the crystal structure becomes that of the half-Heusler (HH) compound ($C1_b$ type, space group $F\bar{4}3m$), as shown in Figure 1. FH and HH compounds are known for their multifunctional properties that include semiconductor bandgap, half-metallic ferromagnetism, superconductivity, topological surface states, piezoelectricity, and shape memory effect just to name a few [17].

Half-Heusler compounds have been synthesized by many research groups worldwide. Sample fabrication is relatively uncomplicated and production in kilogram quantity has been demonstrated[19]. Since the late 1990s, half-Heusler alloys have garnered much attention as prospective thermoelectric materials for mid-to-high temperature energy conversion view of their high Seebeck coefficient, moderate electrical resistivity, and high thermal stability[20, 21, 22, 23, 24, 25]. The main disadvantage of these materials is their relatively high lattice thermal conductivity. Nevertheless, as demonstrated by many groups, atomic mass and atomic size fluctuations and various kinds of defects due to alloying could enhance phonon scattering, resulting in reduction of lattice thermal conductivity. Another advantage of HH compound is that each of the three sublattices can be independently doped. Thus, ideally the power factor and thermal conductivity can be tuned separately. In view of the various material advantages mentioned, it is not surprising that half-Heusler compounds have emerged as one of the most studied TE materials in recent years[26, 27, 28].

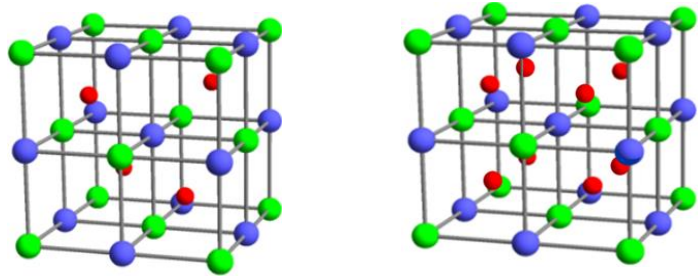

Figure 1. Unit cells of half-Heusler compound (left) and full-Heusler compound (right). Symbols: X(●), Y(●), and Z(●). Starting from the XY zinc blend structure, HH compound can be described as a "half-stuffed zinc blend" and FH compound a "fully-stuffed zinc blend".

*2.3 State-of-the-art thermoelectric materials in the mid-to-high temperature range*

ZT versus *T* plots for various thermoelectric materials can be found in several review articles referenced herein. In view of recent progress made on half-Heusler compounds and their potential roles in mid-to-high temperature energy conversion, it is timely to provide an up-to-date comparison of these materials with other state-of-the-art high ZT compounds, particularly chalcogenides and skutterudites. The plots in Figure 2 are made following the format of figure 6 in reference 29. The competitiveness of half-Heusler compounds, in terms of thermoelectric performance and materials stability and scalability, will undoubtedly continue to generate a plethora of research activities due to the need for practical TE materials.

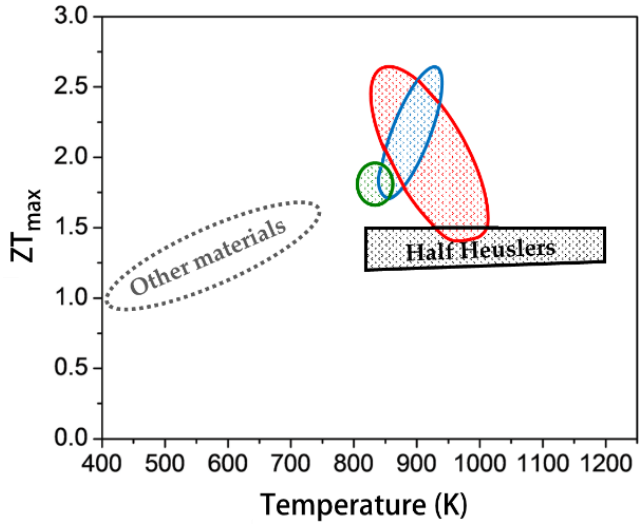

Figure 2. Highest ZT ($ZT_{max}$) and temperature at which $ZT_{max}$ occurs is shown for four state-of-the-art mid-to-high temperature thermoelectric compounds: Cu-based chalcogenides (red), skutterudites (green), Sn- and Pb-based chalcogenides (blue), and half-Heuslers (black). The half-Heuslers in the shaded region are those of Gen-3 HH materials presented in Section 5.2.

*2.4 Thermoelectric conversion efficiency*

In its simplest form, the ideal maximum conversion efficiency η for a thermoelectric device or generator (TEG) is given by the following expression[27]

$$\eta = \left(\frac{T_h - T_c}{T_h}\right)\left(\frac{\sqrt{1+\overline{ZT}}-1}{\sqrt{1+\overline{ZT}}+\frac{T_c}{T_h}}\right) \quad (4)$$

where the first term in parenthesis on the right is the Carnot efficiency, $T_h$ is the hot-side temperature, $T_c$ is the cold-side temperature, and $\overline{ZT}$ is the temperature-averaged ZT. In reality, η(T) is determined by the temperature-dependent TE parameters and not by the average ZT. The relevant equations for calculating η(T), which can be found in ref. 30, are highlighted in Figure 3. If the n-leg and p-leg have the same ZT, our expression for η converges to the results of Kim *et al*[31]. In the absence of unusual temperature dependence, the η values calculated using equation (4) and those shown in Figure 1 differ only by a few percent.

The ideal TE conversion efficiencies calculated for several representative state-of-the-art thermoelectric materials are plotted in Figure 4. The few points plotted represent the highest η(T) values for these TE materials based on the ZT data obtained from the numerous sources cited in ref. 27 xx (see Figure 2 and caption shown therein) and ref. 26 xx (see Table 1 shown therein), as well as most recent results for high ZT half-Heuslers discussed in Sections 5.2 and 5.3. The TE data for α-MgSbAg can be found in Ref. 32. xx From the plot, the half-Heusler compounds are seen to have high conversion efficiency comparable to those of chalcognides and skutterudites. Despite the recent advancement in ZT, the present TEG conversion efficiencies lag behind those of a Rankine steam engine or Stirling gas engine. There are areas of waste heat recovery where research and applications of thermoelectric conversion technologies are still active. The interested reader is encouraged to read the commentary by Vining[33].

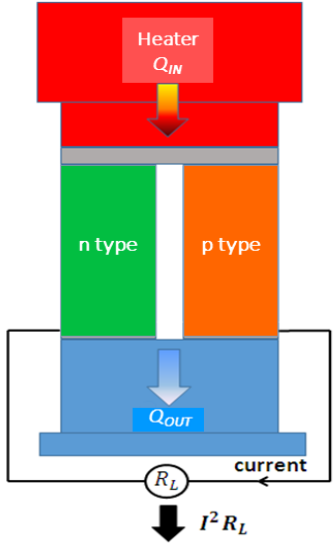

$$\eta = \frac{\text{output power}}{\text{heat input to the generator}}$$

$\eta$ is maximum conversion efficiency obtained by optimizing load resistance $R_L$ and ratio of base areas of n and p legs

$$\eta = \frac{\Delta T}{T_h} * \frac{\left(1 + \Delta T * \left(\frac{\overline{S}}{(\overline{\rho_n}*\overline{k_n})^{\frac{1}{2}} + (\overline{\rho_p}*\overline{k_p})^{\frac{1}{2}}}\right)^2 \left(\frac{S_h * T_h}{\overline{S}*\Delta T} - \frac{1}{2}\right)\right)^{\frac{1}{2}} - 1}{\frac{S_h}{\overline{S}}\left((1 + \Delta T * \left(\frac{\overline{S}}{(\overline{\rho_n}*\overline{k_n})^{\frac{1}{2}} + (\overline{\rho_p}*\overline{k_p})^{\frac{1}{2}}}\right)^2 \left(\frac{S_h * T_h}{\overline{S}*\Delta T} - \frac{1}{2}\right))^{\frac{1}{2}} + 1\right) - \frac{\Delta T}{T_h}}$$

$\Delta T = T_h - T_c$

$\overline{S} = \frac{1}{\Delta T}\int_{T_c}^{T_h}\left(S_p(T) - S_n(T)\right)dT$       $S_h$ = Seebeck coefficient at $T_h$

$\overline{k_n} = \frac{1}{(T_h - T_c)}\int_{T_c}^{T_h}(k_n(T))dT$       $\overline{k_p} = \frac{1}{(T_h - T_c)}\int_{T_c}^{T_h}(k_p(T))dT$

$\overline{\rho_n} = \frac{1}{(T_h - T_c)}\int_{T_c}^{T_h}(\rho_n(T))dT$       $\overline{\rho_p} = \frac{1}{(T_h - T_c)}\int_{T_c}^{T_h}(\rho_p(T))dT$

Figure 3. (Left) Schematic sketch of a thermoelectric n-p module. The symbols are self-explanatory. (Right) Equations for computing the maximum conversion efficiency. $S_h$ is Seebeck coefficient at the hot end. Thermal conductivity is labeled $k$ following reference 30.

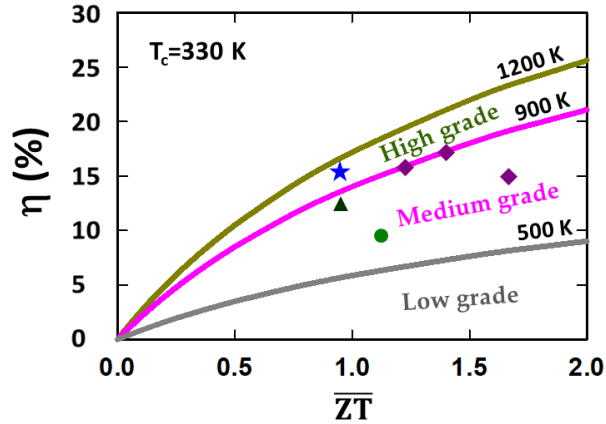

Figure 4. η (in percent) versus $\overline{ZT}$ plots obtained using equation (3). The cool-side temperature $T_c$ is set at 57°C/330K. The three isotherms shown approximately partition the three temperature ranges of waste heat sources mentioned in Section 1. The corresponding Carnot efficiencies at 500, 900, and 1200K are 34, 63, and 72%, respectively. η for several state-of-the-art compounds with highest $\overline{ZT}$ are plotted. These compounds include Pb-, Sn-, and Cu-based chalcogenides (solid diamond), half-Heuslers (solid star), skutterudites (solid triangle), and α-MgSbAg (filled circle) with peak ZT~2.3-2.7, 1.5, and 1.4, respectively. The corresponding $\overline{ZT}$ are ~1.25-1.7, 0.94, 0.97, and 1.1, respectively.

# 3. Half-Heusler Compounds

*3.1 Electronic structure*

The electronic structure and structural stability of half-Heusler compounds have been discussed quite extensively in the literature over the last two decades. Several recent papers on high-throughput screening of stable HH and FH compositions are recommended for reading[13, 14, 15, 16, 18]. A half-Heusler compound is semiconducting when the electron valence count (VEC) is 8 (e.g. LiMgP[34], bandgap ~2.4 eV) or 18 (e.g HfNiSn[9] and TiCoSb[35, 36], bandgap ~0.5 and 0.9 eV, respectively) Those with VEC away from 8 and 18 are metallic. Band structure calculations using FLAPW (full linearized augmented plane wave method)[25] and LMTO-ASA (linearized muffin tin orbitals - atomic sphere approximation)[37] showed that hybridization between the d orbitals of Zr and Ni led to the formation of the bandgap (d-d gap). Moreover, the electronic bands of ZrNiSn directly below the gap have strong Ni-d character, while those above the gap are of primarily Zr-d character. Sn-p levels contribute little to the density of states (DOS) in the vicinity of the band gap, indicating that doping that site would merely result in increasing or decreasing the valence electron count without otherwise modifying the band structure near the Fermi level $E_F$. Below the gap lies four Sn/Sb s and p bands and five bands of predominately Ni- or Co-d character. When VEC=18, the nine bands are filled and all the bonding states are then occupied while all antibonding or non-bonding states are empty[38], giving rise to semiconducting behavior. The role of atomic d-state level in bandgap structure was validated by electrical transport measurement[39, 40]. The measurements suggested that introduction of transition metal dopants resulted in the formation of dopant bands within the gap in positions depending on the valance d-shell filling of the dopant atomic species. The observed trend can be understood in terms of the energy level trend of the d states in the dopant elements[39, 41]. It is worth mentioning that bandgap formation in half-metallic half-Heuslers with VEC>18, an important class of spintronic materials, have also been studied quite extensively. The total spin moments (M) in HH half-metals followed a Slater-Pauling behavior encapsulated in the equation $M = VEC - 18$ [15, 17, 42].

Transition metal and refractory metal based half-Heusler semiconducting compounds have bandgap in the range of a few tenths of eV, or several times 1000K on the thermal scale. The bandgap size is most suited for thermoelectric applications in the mid-to-high temperature range. Together with the various material advantages mentioned, namely high power factor, good thermal stability, and demonstrated scalability, it has invigorated the study of half-Heusler compounds for mid-to-high temperature thermoelectric conversion [26, 27, 28].

*3.2 Synthesis methods*

The various methods for synthesizing half-Heusler compounds have been highlighted in two recent review articles[43, 44]. Many of the half-Heusler compounds studied, particularly those that contained refractory metals with high melting points, were usually produced in the form of ingots using one of the high temperature alloying methods. The latter methods include arc melting, induction melting, and levitation melting. In addition, optical floating zone method and microwave heating have also been used to synthesize HH compounds. The ingots produced from arc melting and induction melting usually contained secondary phases in addition to the HH phase. The samples were typically annealed at ~650-850°C for 7-10 days to produce single-phase samples [20,

[21, 22, 45, 46]. On the other hand, single-phase ingots could be synthesized using levitation melting[47, 48]. The samples produced from the methods mentioned inevitably showed various degrees of porosity. Starting in 2005- 2006, several groups in Japan began to utilize spark plasma sintering (SPS) technique to produce HH samples for thermoelectric measurement[49, 50, 51]. In comparison with as-cast samples, the SPSed samples were essentially free of porosity and therefore more robust mechanically. Incidentally, it was one of these groups that first reported a high ZT of ~1.5 at 700K in some SPSed n-type (Hf,Zr)TiNiSn samples[51].

The composition of HH compounds must be carefully determined in order optimize TE properties. Especially in the case of minor alloying such as doping, a small change in the composition can have a drastic effect on the results. Thus, one must be able to control the alloy composition with considerable precision. The dopant content reported so far was at most a few atomic percent or less. As an example, for Sb and V doped (Hf,Zr)NiSn compounds, the author's laboratory first made the $Sn_{90}Sb_{10}$ or $Hf_{90}V_{10}$ precursor ingot. A small amount of extra Sb was added in order to compensate for some loss of Sb due to evaporation[52]. The HH ingot was then produced by arc melting proper amounts of elemental Hf, Zr, Ni, Sn and chunks cut from the pre-melted Sn-Sb or Hf-V ingot. The composition and compositional homogeneity were ascertained by Energy-dispersive spectroscopy (EDS) measurement [52, 53].

*3.3  Initial studies of thermoelectric properties*

As mentioned, half-Heusler compounds began to gain attention of thermoelectric researchers in the late 1990s. Earlier, Cook et al[7] reported a power factor of ~2.1 W/m-K at 750K in as-grown TiNiSn samples. Although the high thermal conductivity of TiNiSn resulted in ZT of only 0.3 [22], the potential of half-Heusler compounds as thermoelectric materials was duly recognized. The studies that followed were focused on the dopability of these compounds and the effectiveness of the dopants, as well as the effect of alloying on thermal conductivity [20, 21, 22, 23, 24]. Most of the measurements were conducted up to 300K. The results from these studies were discussed in a review article by the present author[46]. Large Seebeck coefficients approximately in the range of 100-400 µV/K (i.e. in magnitude) and thermal conductivity approximately in the range of 3-20 W/m-K were reported by various groups in a variety of HH compositions. The highest room-temperature power factor (PF) for the undoped compounds was measured on TiNiSn (0.22 W/m-K)[22] and ErNiSb (0.26 W/m-K)[54], while the lowest thermal conductivity (κ) was found in (Zr,Hf)(Co,Pt)(Sn,Sb) (3.1 W/m-K)[55] and TmNiSb (2.8 W/m-K)[54]. The transition metal based HH compounds studied tended to show opposite trends in PF and κ, that is, those that had higher PF tended to have higher κ and vice versa. Doping was employed to enhance PF without increasing κ at high temperature. A high power factor of 3.85 W/m-K as well as ZT of 0.45 was obtained in $(Ti_{0.5}Hf_{0.5})Ni(Sn_{0.975}Sb_{0.025})$ at 700K[46].

*3.4  First generation (Gen-1) thermoelectric half-Heusler materials*

As mentioned in Section 1, the layout of this article is organized according to the three generations (abbreviated Gen-1, Gen-2, Gen-3) of TE half-Heusler materials. A high-level summary of these materials is shown in Table 1. In this section, we will highlight some of the high-performance Gen-1 half-Heusler materials reported in the 2000's. Gen-2 and Gen-3 HH materials will be featured in Section 4 and Section 5, respectively. Among Gen-1 HH materials, the highest ZT was reported to be ~0.85 for n-type $Hf_{0.75}Zr_{0.25}NiSn_{0.975}Sb_{0.025}$[56, 57] and ~0.5 for p-type $Zr_{0.5}Hf_{0.5}CoSb_{0.8}Sn_{0.2}$[58], both occurring at 1000K. The higher ZT compared with those of unalloyed ternary HH compounds was achieved through alloy substitution that increased phonon scattering and power factor, as well as by doping that enhanced the power factor. Earlier, an international team obtained ZT~0.7 at 800K for $Zr_{0.5}Hf_{0.5}Ni_{0.8}Pd_{0.2}Sn_{0.99}Sb_{0.01}$ synthesized using SPS before the latter synthesis technique was widely adopted in the US[59]. Besides the beneficial effect of doping mentioned above, mass fluctuations and strain field fluctuations due to differences in atomic sizes and interatomic force couplings resulted in point defect scattering that significantly reduced $\kappa$[60]. Most recently, Liu et al[61] ascribed their measurement of ZT~1 in Nb-doped (Zr,Hf)CoSb compounds to the mass fluctuation and "lanthanide contraction" effects. Lanthanide contraction resulted in the small atomic size difference between Zr and Hf. The similar atomic size ensured a relatively uniform atomic potential profile, and therefore did not affect the carrier mobility. Meanwhile, the difference between atomic mass of Zr and Hf gives rise to mass fluctuation, suppressing the thermal conductivity.

Table 1. Chronological classification of half-Heusler *compounds* (interchangeable with *materials* to include composites) based on advances in synthesis methods and technical approaches as well as increase in ZT. ZT values are cited in the text. The widely adopted term "spark plasma sintering" is used herein even though "current-assisted sintering" is technically more accurate. The dawn of the Gen-3 era, approximately, overlapped with the publication of two review articles on thermoelectric HH compounds [44, 44].

| HH materials classification | Synthesis | Technical approaches | Effects | Highest ZT |
|---|---|---|---|---|
| Gen-1 (2000's) | almost exclusively melting and annealing | alloying, doping | alloy scattering lowers $\kappa_L$, doping optimizes PF | 0.85 (n) 0.5 (p) |
| Gen-2 (late 2000's to early 2010's) | widely adopted spark plasma sintering and hot pressing | sample densification, nanostructuring | robust samples, grain refinement reduces $\kappa_L$, enhancing S | ~1 |
| Gen-3 (post early 2010's) | combination of mentioned methods | phase separation, structure ordering, band engineering | additional mechanisms (see section *5.2*) | ~1.5 |

## 4. Nanostructured Half-Heusler Materials

*4.1 Microstructural features*

There are principally two types of nanostructured half-Heusler materials, namely: (i) nano-bulk materials composed nearly entirely of nano-size grains, and (ii) composite materials that contain submicron or micron size particles or domains. Domains are usually associated with phase separation. As discussed extensively in the review article by Xie et al [43], and Chen and Ren[44] and references therein, nano-bulk HH materials were fabricated by consolidating micron-size particles consisting of nano-size grains using spark plasma sintering or hot pressing. The latter thermo-mechanical processes gave better control of the microstructure through grain growth or grain refinement, and thus were suitable for producing nano-bulk materials. The process usually occurred at ~80% of the melting temperature under an applied pressure of 50-1000 MPa. The powder particles were produced either by mechanical alloying of elemental powders, melt spinning of molten alloy, or mechanical attrition of the pulverized alloy ingot. As reviewed by Xie et al, nanocomposites were formed either via in-situ growth or ex-situ addition. Figure 5 shows the sketches of three kinds of composite microstructures.

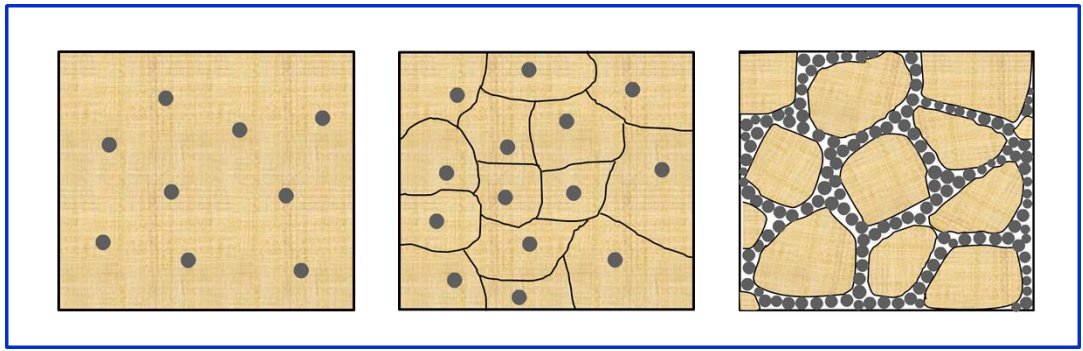

Figure 5. Left panel: Nanoparticles in monocrystalline host. Middle panel: Nanoparticles in polygrained host. Right panel: Nanoparticles aggregated at grain boundaries. The composites studied usually contain different sizes of particles or domains that are more efficient at scattering phonons. A nano-bulk phase is formed when the sample volume is filled with nanoparticles.

*Spontaneous growth of nano-bulk HH materials via recrystallization* – Mechanical milling has been commonly utilized to produce half-Heusler nanograins with size less than 100 nm. The nano-size grain structure was found to be retained in nano-bulk half-Heusler samples produced by shock compaction[62]. The nano-bulk structure was confirmed by atomic force microscopy (AFM) study. Figure 6 shows the AFM micrographs. The average grain size was calculated to be ~35 nm with a standard deviation of ~6 nm. On the other hand, SPS and hot pressing of ball-milled powders often resulted in grain growth in the grain size increasing to ~100-400 nm[63, 64]. Recently, it was reported that the nano-bulk state could also be realized via a recrystallization process[53]. This was observed in some SPSed $Hf_{0.6}Zr_{0.4}NiSn_{0.995}Sb_{0.005}$ nano-bulk samples. The original arc-melted ingot contained micron-size (Hf,Zr)Ni$_2$Sn full-Heusler phase and two HfZr intermetallic phases with Ni and Sn, respectively, in addition to the HH phase. The ingot was pulverized prior to compaction by SPS. After SPS, a single-phase nano-bulk sample was obtained with grain size in the range 50-200 nm. Recrystallization apparently occurred during the transformation from the mixed phase state to single phase state that led to the nano-bulk state. The microstructures of the investigated $Hf_{0.6}Zr_{0.4}NiSn_{0.995}Sb_{0.005}$ samples in the arc-melted and annealed, SPSed, and post-SPS annealed states are shown in Figure 7.

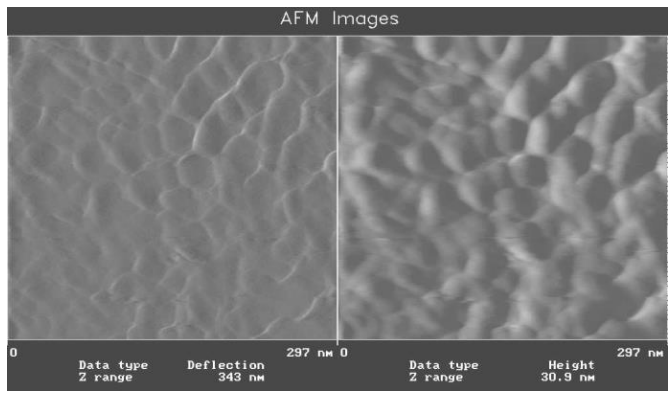

Figure 6. AFM images obtained in the deflection mode (left) and height mode (right) on shock-consolidated TiNiSn(Sb) powder compact (Image: 297nm x 297nm). Preparation of alloy powders: ingot was pre-annealed at 900ºC for 14 hours prior to ball milling.

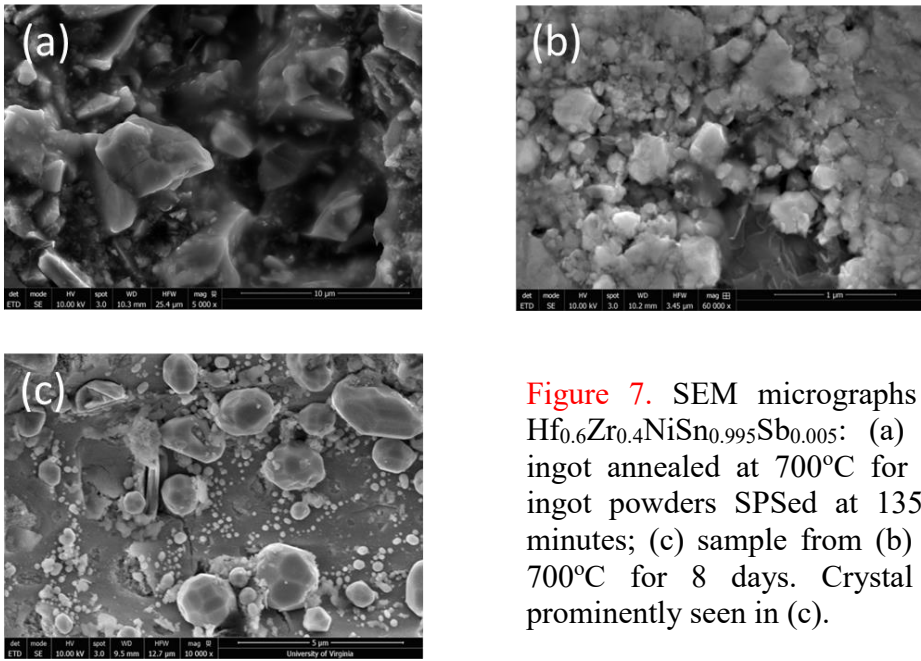

Figure 7. SEM micrographs for n-type $Hf_{0.6}Zr_{0.4}NiSn_{0.995}Sb_{0.005}$: (a) arc melted ingot annealed at 700°C for 8 days; (b) ingot powders SPSed at 1350°C for 30 minutes; (c) sample from (b) annealed at 700°C for 8 days. Crystal facets are prominently seen in (c).

*Nano- and micro-structures from phase separation* – In-situ formation of nanostructured HH composites occurred in directional grown $Ti_{0.37}Zr_{0.37}Hf_{0.26}NiSn$[65], annealed ingot of $Ti_{0.5}Hf_{0.5}CoSb_{0.8}Sn_{0.2}$[66], and annealed cold-pressed[67] and SPSed[68] $Ti_{0.5}Zr_{0.25}Hf_{0.25}NiSn$ powders. The obtained samples showed a two-phase microstructure characterized by nano-micro scale Ti-rich and Hf-rich regions in the matrix. It is evident that (Ti,Hf,Zr)-based HH compounds are prone to phase separation, which is likely due to the smaller atomic size of Ti compared with Hf and Zr. The heterogeneous microstructure in phase-separated materials enhanced phonon scattering, reducing the lattice thermal conductivity, which led to high ZT in some Gen-2 HH materials (Table 2). On the other hand, phase separation was also reported for (Hf,Zr)NiSn using high-resolution synchrotron radiation powder x-ray diffraction[69]. In addition to experimental studies, phase separation in (Ti,Zr,Hf)-based HH alloys was also confirmed in combined ab initio/Calphad computation of binodal and spinodal isopleths for quaternary (Ti,Hf)NiSn and (Ti,Zr)NiSn by

Gurth and Rogl et al[70]. The latter authors pointed out that binodal and spinodal decompositions could be avoided by sintering/annealing the samples at above the critical temperatures. Separately, Berche et al performed ab initio calculation that also revealed phase separation in the (Ti,Zr,Hf)-based HH alloys in agreement with Calphad results[71]. An important message from the studies mentioned is that one may design heat treatment protocol to control the microstructure in (Ti,Zr,Hf)-based HH compounds as a practical approach to improving thermoelectric properties.

*4.2 Second generation (Gen-2) half-Heusler materials, ZT~1.*

Since 2010, the popular utilization of thermo-mechanical processing methods based on spark plasma sintering and hot pressing to synthesize dense samples (higher than 95% of the theoretical density) have led to ZT near 1, or higher, for both n-type and p-type half-Heusler materials [43, 44]. It should be pointed out that ZT near 1 was also reported for samples synthesized using conventional growth technique[65, 66]. The peak ZT in n-type and p-type compounds tended to occur in the temperature regions of 700-900K and 950-1050K, respectively. The improvement of TE properties and associated ZT were ascribed to enhanced scattering of charge carriers and phonons due to presence of nano- and micro-scale structures[63, 64, 72, 73, 74, 75, 76, 77, 78, 79] produced by the various nanostructuring methods mentioned in the previous Section. On the other hand, the enhancement can also be attributed to grain refinement[47, **Error! Bookmark not defined.**] or phase separation as revealed in microscopy studies[65]. Thus, ZT enhancement can also occur without the need for ball milling or melt spinning. Grain refinement in SPSed samples, as discussed in the previous section, deserves further investigation. Besides ball milling, hierarchical microstructure, which could exist in the heterogeneous samples mentioned, is ideal for scattering both short and long wavelength phonons in reducing lattice thermal conductivity[73, 80, 81]. Charge carriers can also be trapped at the interface, providing a potential barrier to enhance the Seeback coefficient[82, 83, 84, 85, 86, 87]. Such enhancements were observed in nanostructured half-Heusler materials[63, 72, 88, 89, 90]. Some examples are shown in Figure 8. The synthesis methods employed to produce Gen-2 materials and their microstructural features are highlighted in Table 2.

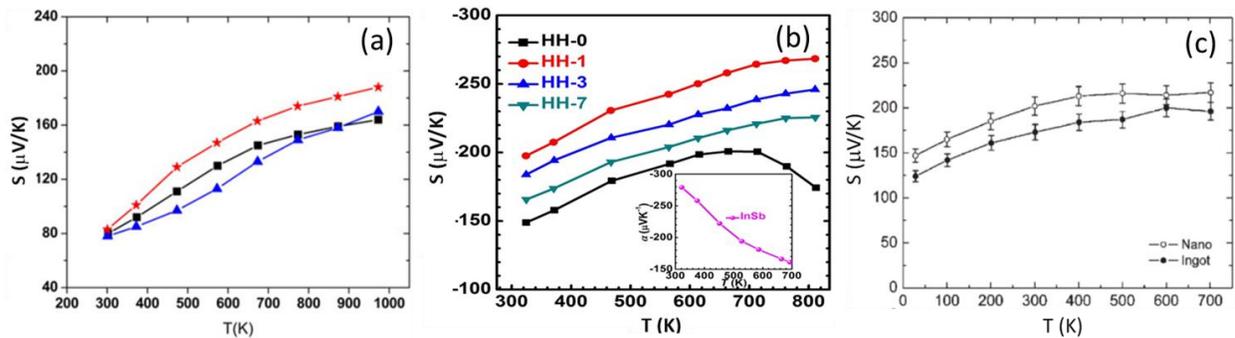

Figure 8. Thermopower enhancement in nanostructured half-Heusler materials: (a) $Hf_{0.3}Zr_{0.7}CoSn_{0.3}Sb_{0.7}$ dispersed with 0 (square), 1(triangle), and 2 vol. % (star) $ZrO_2$ nanoparticles (ref. 72); (b) $Ti_{0.5}Zr_{0.25}Hf_{0.25}Co_{0.95}Ni_{0.05}Sb$ with 0, 1, 3 and 7 at. % of InSb nanoinclusions (ref. 88); (c) $Zr_{0.5}Hf_{0.5}CoSb_{0.8}Sn_{0.2}$ nano-bulk and ingot samples (ref. 63). Improvement in ZT of these three systems comes from the simultaneous increase in S and decrease in κ.

Table 2. Listed are representative Gen-2 half-Heusler materials (both n-type and p-type) with ZT~1, methods of synthesis, and microstructural features. Abbreviations: Ball milling (BM), hot press (HP), spark plasma sintering (SPS), arc melting (AM), levitation melting (LM), and melt spinning (MS).

| Representative HH materials | Synthesis methods | Microstructural features | Type (n or p) | Refs. |
|---|---|---|---|---|
| (Hf,Zr,Ti)Co(Sb,Sn) | BM and HP | nanograined structure | p | 74 |
| Sb-doped (Hf,Zr,Ti)NiSn | BM and HP | nanograined structure | n | 75 |
| (Ti,Zr,Hf)NiSn | Directional growth | phase separation | n | 65 |
| (Ti,Zr,Hf)Co(Sb,Sn) | AM and annealed | phase separation dendrites | p | 66 |
| Sb-doped (Hf,Zr)NiSn | LM and SPS/HP | grain refinement | n | 47 |
| Sb-doped (Hf,Zr)NiSn | MS and SPS | nanograins embedded in submicron grains | n | 73 |
| (Hf,Zr)Co(Sn,Sb)/ nano-$ZrO_2$ | AM and SPS | embedded nanoparticles | p | 72 |
| Sb-doped (Hf,Zr)NiSn | AM and SPS | nano- and micron-grained structures | n | 72 |

*4.3 Conversion efficiency of TEG built with Gen-2 half-Heusler materials*

The power conversion efficiency ($\eta$) of single-couple thermoelectric generator (TEG) built with Gen-2 half-Heusler materials, specially n-type $Hf_{0.6}Zr_{0.4}NiSn_{0.995}Sb_{0.005}$ and p-type $Hf_{0.3}Zr_{0.7}CoSn_{0.3}Sb_{0.7}/ZrO_2$, was found to reach ~9 % near 1000K[72]. The TEG delivered a power density near 8.9 W/cm$^2$, decreasing to ~3.5 W/cm$^2$ for a 49-couple TEG (unpublished results). The $\eta$ achieved was near $\eta_{ideal}$ ~ 9.2% using the expressions in Section 2.3. The results are shown in Figure 9. For comparison, typical $\eta$ for commercial TEGs is 5-8%. Cook et al reported conversion efficiency of 20% for a three-stage cascade TEGs[91]. The assembly of the cascade devices included nano-bulk $Bi_2Te_3$ materials on the cold-side bottom stage, PbTe and TAGS ($AgSbTe_2$-GeTe) compounds for the middle stage, and half-Heusler materials for the high-temperature top stage. Figure 10 shows the schematic diagram of the cascade device and the conversion efficiencies obtained in several tests of the device performance. The high TE conversion efficiency achieved in Gen-2 HH materials indicated that these materials are promising candidates for mid-to-high temperature waste heat recovery applications. Compared with the half-Heuslers, other prospective TE materials such as skutterudites and Pb-based chalcogenides have issues with poor thermal stability, medium to high toxicity, and weak mechanical strength.

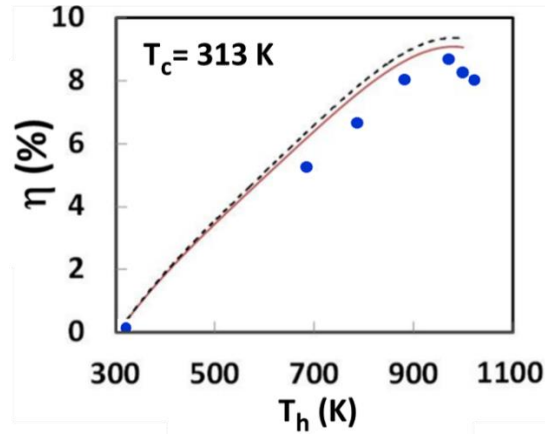

Figure 9. Thermoelectric power conversion efficiency for a n-p couple module built with n-type $Hf_{0.6}Zr_{0.4}NiSn_{0.995}Sb_{0.005}$ compound and p-type $Hf_{0.3}Zr_{0.7}CoSn_{0.3}Sb_{0.7}/ZrO_2$ nanocomposite. The plots shown are: experimental data (blue filled circle), calculated results using temperature-averaged ZT (dashed line), calculated results using temperature-dependent TE properties (solid line) discussed in section 2.4. Figure is adapted from reference 30 with some modification.

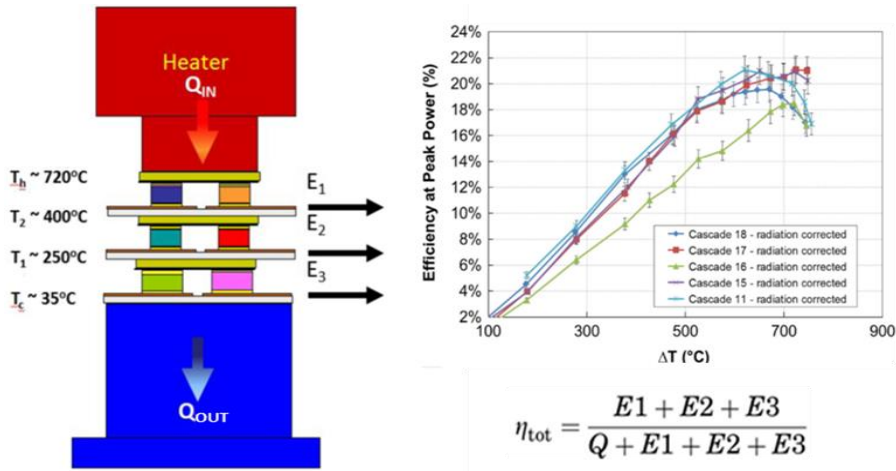

Figure 10. (Left) Schematic of a 3-stage cascade TEG. (Right) Summary of cascade efficiencies obtained in power testing. $\eta_{tot}$ is the total conversion efficiency of the cascade. The content of the figure is extracted from Ref. 91. The schematic TEG is slightly modified for clarity purpose.

## 5. Recent Development of Half-Heusler Materials

The electronic and vibrational contributions to thermoelectric properties are intertwined (c/o Section 2). Nanostructure influences both thermal conductivity and thermopower. Alloy composition influences both thermal conductivity and band structure. At the fundamental level, such complexity appears to suggest a departure from the "phonon glass electron crystal" paradigm. Nonetheless, recent efforts by several groups have succeeded in optimizing the ZT in half-Heusler compounds. In this section, we will focus on the advances made in thermoelectric half-Heusler compounds in the past few years. Two prior review articles[43, 44] on thermoelectric HH compounds have provided us with an excellent account of the state-of-the-art until 2013.

*5.1 Third generation (Gen-3) half-Heusler materials with ZT>1.*

Although nanostructure continued to play an important role in the development of Gen-3 half-Heusler materials, discovery of new HH compositions has emerged as a promising direction for band engineering in improving TE properties. The various improvements achieved have led to the emergence of third-generation (Gen-3) HH materials with ZT reaching 1.5. The highest ZT reported for several n-type and p-type half-Heusler compounds are summarized in Table 3. The origins of high ZT were attributed to various factors, namely, microstructure, mass fluctuations, structural order, dopant resonant states, heavy hole band, and high valence band degeneracy, as well as soft phonons. These various approaches to high ZT in HH compounds are referenced in the below section.

Table 3. Gen-3 half-Heusler materials with ZT>1. Maximum ZT values occur at the temperatures indicated. Included are enabling factors that lead to beneficial effects on thermoelectric properties. Other features may coexist in a less prominent way.

| N-Type Half-Heusler Compounds | | | | |
|---|---|---|---|---|
| Materials | ZT | Enabling recipes | Beneficial effects | Refs. |
| $Ti_{0.5}Zr_{0.25}Hf_{0.25}NiSn$ | 1.2 (830 K) | Phase separation | Lower κ | 67 |
| $Hf_{0.6}Zr_{0.4}NiSn_{0.995}Sb_{0.005}$ | 1.2 (860 K) | Strain reduction | Higher ρ and S, lower κ | 53 |
| $Hf_{0.65}Zr_{0.25}Ti_{0.15}NiSn_{0.995}Sb_{0.005}$ | 1.3 (830 K) | Nano-oxide embedment | Higher ρ and S, lower κ | 30 |
| $Hf_{0.59}Zr_{0.40}V_{0.01}NiSn_{0.995}Sb_{0.005}$ | 1.3 (900 K) | Dopant resonant states | Higher ρ and S, lower κ | 92 |
| $Ti_{0.5}Zr_{0.5}NiSn_{0.98}Sb_{0.02}$ $Ti_{0.5}Zr_{0.25}Hf_{0.25}NiSn_{0.98}Sb_{0.02}$ | 1.2 (820 K) | Presence of Nanograin | Higher ρ and S, lower κ | 70 |
| $Ti_{0.5}Zr_{0.25}Hf_{0.25}NiSn$ | 1.5 (820 K) | Similar to Reference [14] | Higher S, lower ρ and κ | 93 |
| P-Type Half-Heusler Compounds | | | | |
| FeNb$_{0.88}$Hf$_{0.12}$Sb<br>FeNb$_{0.86}$Hf$_{0.14}$Sb | 1.5<br>(1200 K) | Heavy hole band, high dopant content, heavy atomic mass | Lower ρ, low κ, high S | 94 |
| ZrCoBi$_{0.65}$Sb$_{0.15}$Sn$_{0.20}$ | 1.4<br>(973 K) | High hole band degeneracy, high dopant content, low energy phonon | Lower ρ, low κ, high S | 95 |
| Ta$_{0.74}$V$_{0.1}$Ti$_{0.16}$FeSb | 1.52<br>(973 K) | High hole band degeneracy, low energy phonon, point defect | Lower ρ, low κ, high S | 96 |

*5.2 Paths to high ZT beyond nanostructuring*

As Table 3 shows, recent development of thermoelectric half-Heusler compounds involved new approaches that led to beneficial electronic and thermal properties. Band engineering has played a definitive role in achieving ZT as high as 1.5. We will highlight these new approaches and their enabling mechanisms below.

(i) Improve bandgap by reducing lattice strain – Antisite disorder involves transfer of Ni atoms from occupied fcc sublattice to vacant fcc sublattice sites in the HH lattice[97]. Recently, ab initio calculation showed that interstitial defects were also stable in the HH lattice[98]. Structural disorder led to the formation of in-gap states, effectively reducing the size of the bandgap[98, 99]. The structural defects reported could also strain the crystal lattice, which would have an effect on the bandstructure. Investigation of TE properties of disordered HH compounds has recently gained attention[53, 98, 100, 101, 102, 103, 104, 105, 106]. The disorder tends to suppress the lattice thermal conductivity, resulting in enhanced ZT. In one of the studies, it was found that the strain in HH compounds could be reduced if the samples were sintered at sufficiently high temperatures just below melting[53]. The X-ray pattern for Hf$_{0.6}$Zr$_{0.4}$NiSn$_{0.995}$Sb$_{0.005}$ is shown in Figure 11 (a). The inset shows peak widths of the (220) reflection obtained on samples that were synthesized under different conditions. As Figure 11 (b) shows, the strain decreases with increasing synthesis temperature, from samples annealed at 700 °C to those sintered near the melting point (~1450 °C), by as much as 30%. The effect of strain on thermoelectric properties is notable, as can be seen in the large enhancement in the power-factor temperature product, PF*T. The latter increased by nearly 50% as the synthesis temperature increased from 700ºC to 1350ºC. The results are shown in Figure 11 (c). This resulted in a 20% increase in ZT, from 1 to 1.2. The increase in power factor, partly responsible for the increase in ZT, was ascribed to the decrease in carrier density and increase in carrier mobility. There was also a corresponding increase in the size of the bandgap. The mitigation of lattice strain has apparently resulted in the reduction of in-gap defect states. The approach to reduce lattice strain to improve TE properties may also be applicable to other thermoelectric materials.

(ii) Enhance thermopower with dopant resonant states – Under the right orbital hybridization condition, resonant states manifest as a localized sub-band in the density of states (DoS) near the Fermi level. The enhanced DoS that resulted can lead to an increase in the Seebeck coefficient, leading to an increase in ZT in some semiconductors[107, 108, 109, 110]. Simonson et al[111] first reported evidence of resonant states by doping half-Heusler compound $Hf_{0.75}Zr_{0.25}NiSn$ with vanadium. The results are highlighted in Figure 12. Carrier density, mobility, effective mass, and magnitude of Seebeck coefficient at room temperature are shown as the concentration of vanadium is increased to near 0.8%. The resonant states scenario is corroborated by the observed correlation between increase in Seebeck coefficient, decrease in carrier concentration, and increase in effective mass. In particular, the correlation between enhanced Seebeck coefficient and increased DoS was confirmed by specific heat measurement.

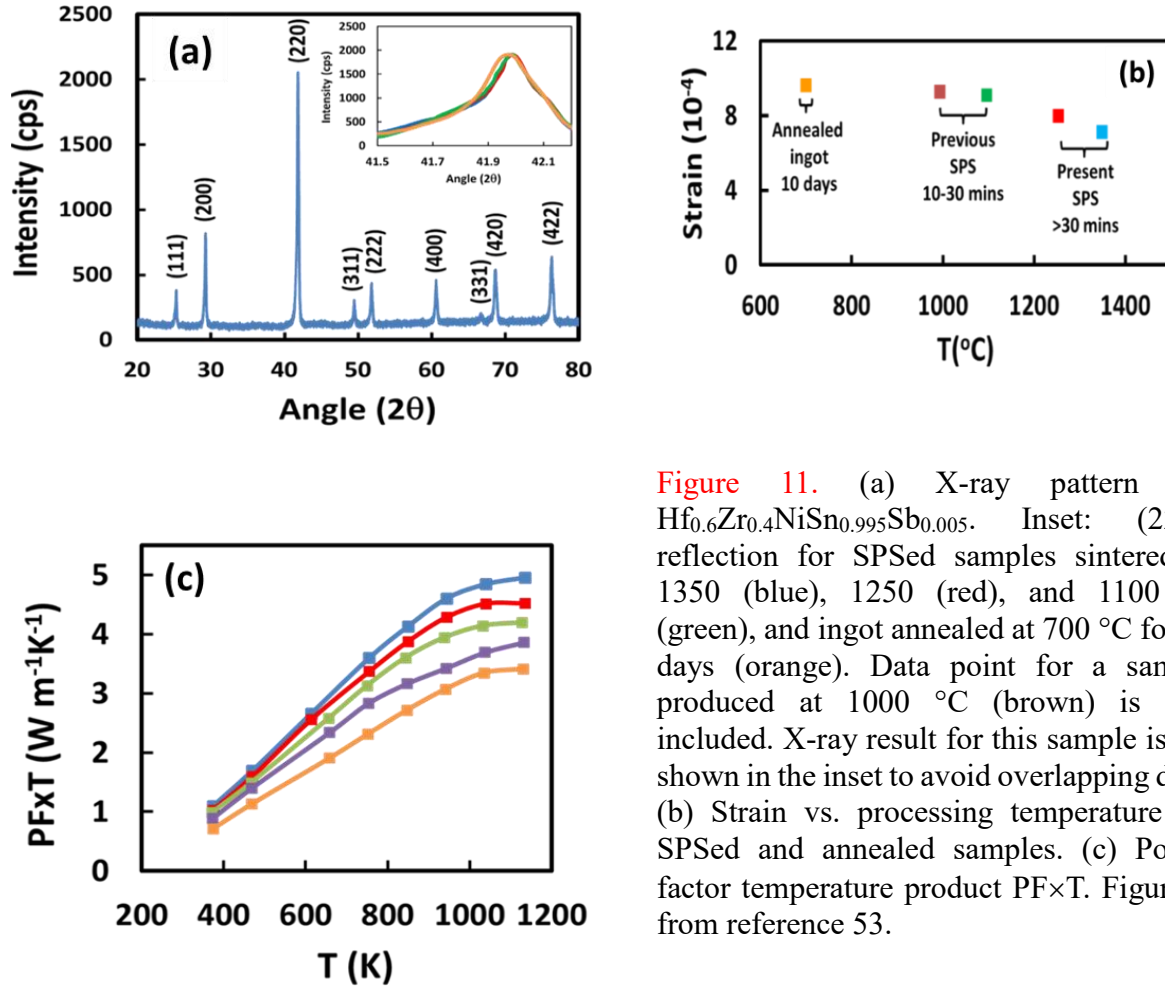

Figure 11. (a) X-ray pattern for $Hf_{0.6}Zr_{0.4}NiSn_{0.995}Sb_{0.005}$. Inset: (220)-reflection for SPSed samples sintered at 1350 (blue), 1250 (red), and 1100 °C (green), and ingot annealed at 700 °C for 10 days (orange). Data point for a sample produced at 1000 °C (brown) is also included. X-ray result for this sample is not shown in the inset to avoid overlapping data. (b) Strain vs. processing temperature for SPSed and annealed samples. (c) Power factor temperature product PF×T. Figure is from reference 53.

Following the finding of Simonson, Chen et al explored improvement of TE properties by doping n-type $Hf_{0.6}Zr_{0.4}NiSn_{0.995}Sb_{0.005}$ that already showed ZT>1 with the VA group elements vanadium, niobium, and tantalum (V, Nb, and Ta, respectively)[92]. Alloys of

$(Hf_{0.6}Zr_{0.4})_{1-x}M_xNiSn_{0.995}Sb_{0.005}$ (M = V, Nb, Ta), where $x$ = 0.002, 0.005, and 0.01 were investigated and their TE properties of these doped samples are shown in Figure 13. V acted as a resonant dopant, while N and Ta acted as regular dopants based on their TE properties measured. Both ρ and S of the V-doped samples were found to increase relative to those of undoped ones, while κ decreased due to the increase in ρ. The power factor of V-doped samples remained the same as the undoped compounds, resulting in a high ZT of 1.3 near 850 K for $(Hf_{0.6}Zr_{0.4})_{0.99}V_{0.01}NiSn_{0.995}Sb_{0.005}$ due to the decrease in κ. In contrast, Nb and Ta doped samples showed an opposite trend, as noted in the decrease in ρ and S, and the power factor also decreased. The increased ZT in V doped samples concomitant with the reduced carrier concentration and increased band mass underscores a band-structure mechanism of ZT enhancement. More recently, beneficial TE transport properties (increased S and decreased κ) and enhancement of ZT, the latter by as much as 70%, was found in V doped ZrNiSn in comparison with the undoped compound[112]. The authors discussed their results in light of V induced resonant states near the Fermi level. The cost effectiveness of Hf-free high ZT half-Heusler compounds was also pointed out.

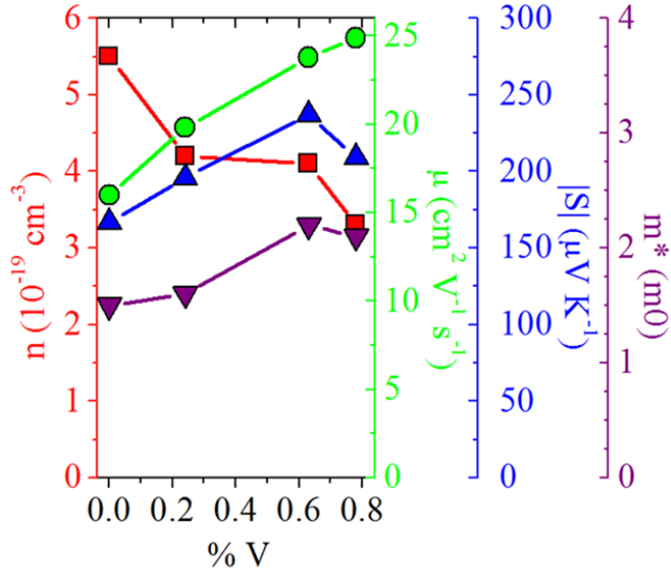

Figure 12: Carrier concentration (first axis, red squares), carrier mobility (second axis, green circles), magnitude of Seebeck coefficient (third axis, blue upward triangles), and carrier effective mass (fourth axis, purple downward triangles) of V-doped $Hf_{0.75}Zr_{0.25}NiSn$ as a function of V concentration. All values were either measured (S) at room temperature or computed (n, μ, m*) from room temperature data. Figure and caption are from reference 111.

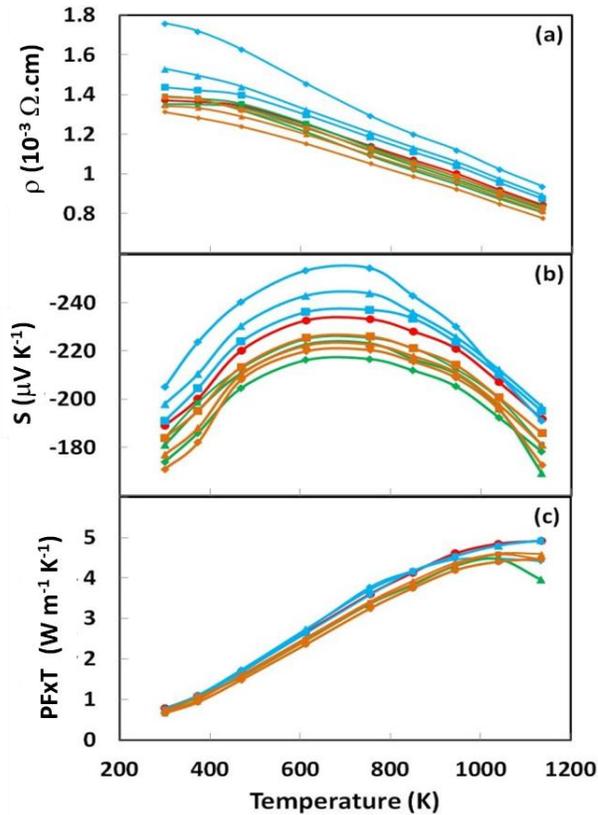

Figure 13. (a) Electrical resistivity, (b) thermopower, and (c) power factor times temperature for M=V, Nb, Ta doped $(Hf_{0.6}Zr_{0.4})_{1-x}M_xNiSn_{0.995}Sb_{0.005}$. $x$ = 0 (red circle); V: $x$ = 0.002 (blue square), 0.005 (blue triangle), 0.01 (blue rhombus); Nb: $x$ = 0.002 (green square), 0.005 (green triangle), 0.01 (green rhombus); Ta: $x$ = 0.002 (orange square), 0.005 (orange triangle), 0.01 (orange rhombus). Overlaps occur in the power factor plots. Figure and captions are from reference 92.

(iii) High ZT via band engineering and soft phonons - In 2015, Fu et al discovered FeNbSb-based p-type half-Heusler compounds with heavy hole band[94, 113]. The authors reported a large effective band mass near 7 to 10 $m_e$ for the new compounds, significantly higher than those reported for other high-performance TE materials. As shown in Figure 14 (a), the high density of states from the narrow hole band enables FeNbSb to accommodate a high doping level. The ultimate benefit to the thermoelectric properties is that the high content of heavy-mass dopant such as Hf optimizes the power factor, simultaneously reducing the lattice thermal conductivity due to the enhanced point-defect scattering. Figure 14 (b) shows a high effective mass, high power factor, and high doping level, all occurring in the same compound. A record high ZT of 1.5 for p-type HH compound was obtained at 1200K upon substituting Nb with 12-14 % of Hf.

FeNbSb based compounds were also studied by other groups. Guo et al observed phase separation in FeTiNbSb, which seems to be a common feature of HH compounds with atomic size mismatch[114]. He et al reported a record high power factor of ~100 μW/cm-$K^2$ (or power factor temperature product of ~3 W/m-K) at room temperature[115]. First principles study of electron-phonon interaction by Zhou et al revealed that the mechanism that led to remarkable high PF in HH compounds lied in the weakness of electron–acoustic deformation potential couplings[116]. These weak couplings originate from crystal symmetry-protected nonbonding orbitals at the band edge. It was emphasized by the authors that the vanishing bonding (antibonding) orbital interactions has made the half-Heuslers unique material platforms that bypass the traditional viewpoint. A more traditional approach to optimize ZT in FeNbSb was studied by Li et al[117].

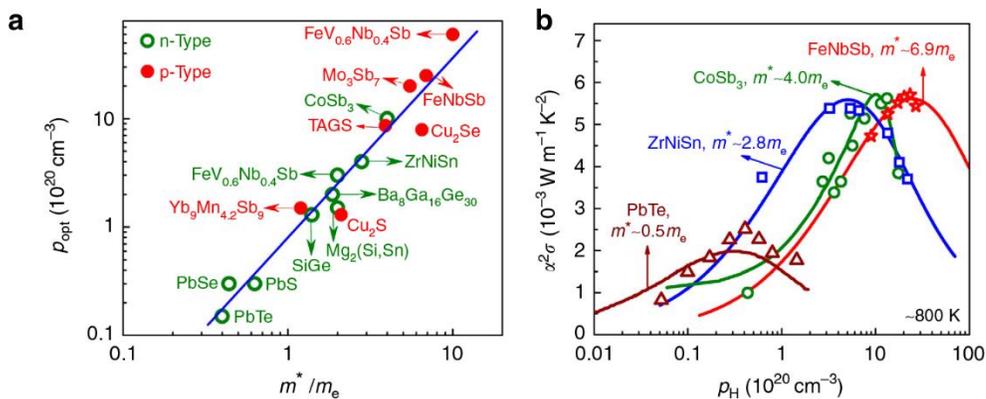

Figure 14. Comparison of transport character of light-band and heavy-band thermoelectric compounds. (a) The optimal carrier concentration $p_{opt}$ is plotted against the density-of-state effective mass m* for different thermoelectric compounds. The solid line is a guide for eyes. (b) Carrier concentration dependence of power factor for the typical light-band PbTe, and the heavy-band system: n-type ZrNiSn, n-type filled $CoSb_3$ and p-type FeNbSb near 800K. Figure and caption are from reference 94.

Recently, Zhu et al discovered new p-type half-Heusler compounds based on ZrCoBi[95] and TaFeSb[96] with record-high ZT of ~1.42 and 1.52 at 973K, respectively. ZT versus T of these compounds is shown in Figure 15. These HH compounds were discovered using the principle of inverse design of materials to search for high-ZT thermoelectric half-Heusler compounds with the required functionality and properties in mind. By adopting this approach, the authors discovered several unreported HH compounds. Among the latter were ZrCoBi and TaFeSb with relatively large bandgaps (~0.5 eV) and high band degeneracy. The outstanding thermoelectric performance was attributed to the unique band structure with a high band degeneracy ($N_v$) of 10 and 8 in the two compounds, respectively. This finding is highlighted in Figure 16. The outstanding performance also benefited from lower mean sound velocity (softer phonons) that resulted in low thermal conductivity. Defect scattering due to additions of elements with lighter atomic mass and atomic size mismatch also contributes to the lowering of $\kappa_L$.

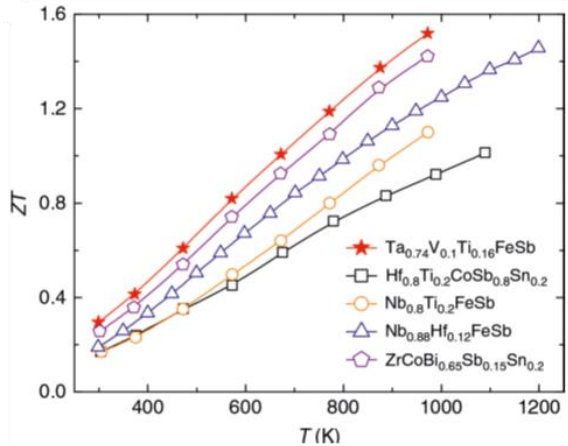

Figure 15. Comparison of the ZT between $Ta_{0.74}V_{0.1}Ti_{0.16}FeSb$ and the other state-of-the-art p-type half-Heuslers. Figure and caption are from reference 96.

*5.3. Power-conversion efficiency of TE modules based on state-of-the-art HH materials*

As mentioned in section 4.2, earlier study reported a power-conversion efficiency η near 9% in TEGs built with Gen-2 HH materials. TEGs based on Gen-3 materials have an ideal conversion efficiency η>10%, and as high as 15%[95, 96]. However, only unileg devices have been tested. η reached 11.4% in one material[95], and power density was 9 W/cm$^2$ in another material[96]. Thus, the power density has not exceeded previously reported value using Gen-2 half-Heusler materials[72]. The fact that η falls significantly below ideal values indicates the need to address outstanding materials science and device fabrication problems.

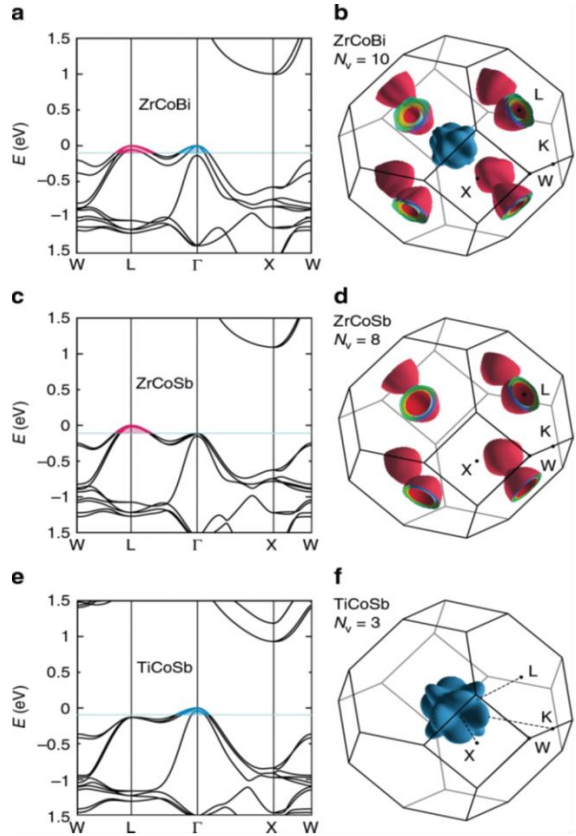

Figure 16. First-principle calculation of band structure. Calculated band structures of ZrCoBi (a), ZrCoSb (c), and TiCoSb (e). The blue lines represent energy level of 0.1eV below VBM. The corresponding iso-energy surfaces at 0.1eV below VBM in Brillouin zone of ZrCoBi (b), ZrCoSb (d), and TiCoSb (f). Figure and caption are from reference 95.

*5.4. Looking ahead*

The vast compositional possibilities for half-Heusler compounds will continue to provide opportunities for optimizing TE properties through design of electronic and phononic structures. Besides the above-mentioned high ZT half-Heusler materials, recent result of ZT~1 in NbCoSb with valence electron count (VEC) of 19 has opened a pathway for searching for new compositions based on the valence balanced principle[118],[119]. The high ZT was found to occur in defect stabilized $Nb_{\sim 0.8}CoSb$, the ground state structure of VEC-19 NbCoSb. Today, various high-throughput methods are utilized in materials screening and design[13, 14, 120]. The predictive capability of these methods will continue to improve as the database expands and new computational algorithms are developed. Despite the positive outlook, important technical issues still need to be overcome in terms of materials transduction. The thermoelectric modules must survive under extended and cyclic mechanical and thermal stresses without the degradation of performance. As such, the half-Heusler materials must have fatigue endurance and some fracture toughness. When coupled with other thermoelectric materials to form a device, the p and n legs must also have compatible linear thermal expansion coefficients. Systematic investigation of the mechanical properties and thermal expansion of refractory half-Heusler compounds was reported by Rogl et al[121]. Fracture toughness of ~2 MPa m$^{1/2}$ and Poisson's ratio near 0.2 were obtained, indicating that refractory half-Heuslers have higher damage tolerance than ceramics. The materials showed linear thermal expansion coefficients comparable to skutterudites and clathrates. Earlier measurement of a p-type refractory half-Heusler compound showed similar elastic modulus and hardness[122].

## 6. Conclusion

Since the initial reports of bandstructure gap and high thermopower in half-Heusler compounds three decades ago, there has been growing interest in these compounds as prospective thermoelectric materials. Half-Heusler compounds show good thermal stability and can be produced in large quantities. Using half-Heusler compounds with only a moderate ZT~1, state-of-the-art conversion efficiency near 9% and high power density output have been demonstrated in p-n couple modules. It is thus not surprising that in recent years, half-Heusler materials have garnered considerable attention in the thermoelectric community as promising TE materials in the intermediate-to-high temperature range. The interest in half-Heusler materials has stimulated new approaches to improve their thermoelectric properties. During the past few years, high ZT>1 reaching ~1.5 have been obtained by targeting structural order/disorder, microstructure, heavy band, high band degeneracy, and dopant resonant states. Thermoelectric conversion efficiency near 11% in some uni-leg demonstration devices has been reported. Despite the impressive progress made to date, however, many technical challenges remain in integrating current high ZT materials into practical devices.